# Theoretical and experimental study of the orientational ordering in the field-induced intermediaite phase from the SmC*$_{FI2}$ phase in chiral smectic liquid crystals

**H. DHAOUADI**[*], **O. RIAHI, R. ZGUEB, F. TRABELSI and T. OTHMAN**

**Université Tunis El-Manar,** *Faculté des Sciences de Tunis, Laboratoire de physique de la matière molle et de modélisation électromagnétique (LP3ME), Campus Universitaire Farhat Hached 2092 Tunis, Tunisie.*

**Abstract**

Under an electric field, chiral smectic liquid crystals transit usually to the unwound SmC* phase where the helical structure is completely unrolled. Sometimes the sample transits initially towards an intermediate polar state before the total destruction of the helix. Based on the extension of the H-T model, a theoretical study of these field-induced phase transitions was carried out. Two hypotheses of the dynamics that give rise to the appearance of the intermediate phase have been discussed. The results of a numerical analysis confirm the known experimental results; the intermediate phase has a three-layer periodicity structure.

[*]Corresponding author. Email: hassen.dhaouadi@ipeit.rnu.tn

## A. Introduction

Ferroelectric liquid crystals are promising materials for fast switching electro-optical displays with wide viewing angle. Incorporating smectic liquid crystals into display devices is extremely attractive. But their widespread commercial use has not yet been realized because of orientation problems and a multitude of structures that appear as a result of a temperature change or by application of an external electric field.

In chiral smectic liquid crystals the molecules are arranged in layers with the director tilted with respect to the layer normal by a temperature-dependent tilt angle **[1].** Many phases are encountered with the same tilt inside the layers but a distribution of the azimuthal direction which is periodic with a unit cell of one (SmC*) two (SmC*$_A$), three (SmC*$_{Fi1}$), four (SmC*$_{FI2}$) or more (SmC*$_\alpha$) layers **[2-4].**

A zero field phase sequence obtained during cooling process is generally in order:

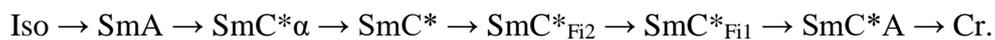

Iso → SmA → SmC*α → SmC* → SmC*$_{Fi2}$ → SmC*$_{Fi1}$ → SmC*A → Cr.

Where Iso and Cr represent the isotropic and crystal phases, respectively. Even so, one or more mesophases may miss in this table but the order of appearance remains the same **[5].** Some theoretical works have shown the possibility of new phases with 5, 6 , 7 … layers periodicity **[6-8].** Liquid crystals are influenced by external effects such as surface anchoring, applied fields and the impurities. This is why we always attend the appearance of new unexpected phases **[9-11].**

New phases can also arise by application of an external electric field. Then it is necessary to draw up the (E,T) phase diagram of these products. It is well-known for a long time that Under an electric field, chiral smectic liquid crystals transit usually to the unwound SmC* phase where the helical structure is completely unrolled **[12].** Sometimes the sample transits initially towards an intermediate polar state before the total destruction of the helix **[13,14].** The dynamics of these transitions was the object of several theoretical and experimental works **[15-19].**

Recently Jaradet et all have discovered, from a study of X-ray resonant diffraction, a new intermediate phase with 3 layers periodicity but different to that of SmC*$_{Fi1}$ phase **.** To interpret their experimental observations they developed a structural model which makes it possible to determine the orientation of the molecules in a layer. They thus showed the evolution of the structure of the *ferrielectric* phase during the increase in the modulus of the applied electric field. **[20].**

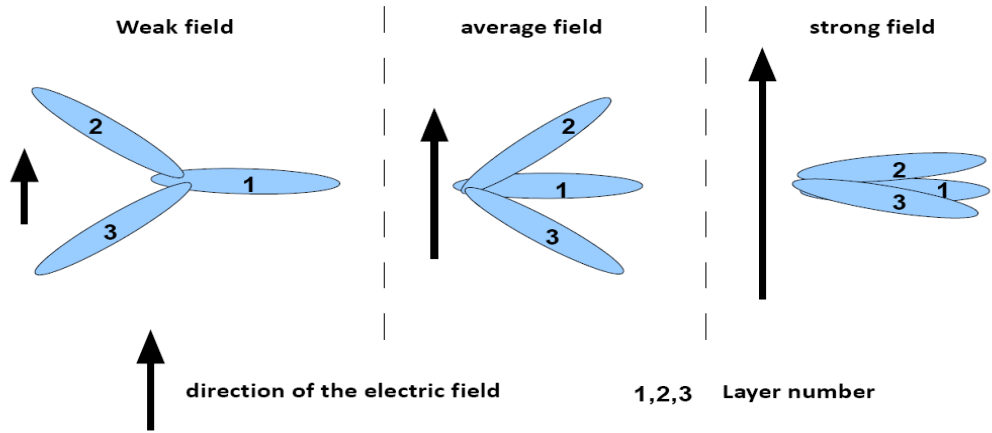

*Figure 1*: *Evolution of the structure of the SmC*$^*_{Fi1}$ *phase as a function of the modulus of the applied electric field.*

In addition, *Helène Gleeson et al* extended their studies to the behaviors under electric field of the three layers *SmC*$^*_{Fi1}$ and four layers *SmC*$^*_{Fi2}$ phases. They present a model for the field-induced phase transitions, based on the writing of the free energies of the studied structures. They showed, at final, the possibility of appearance of a structure noted elementary *Fi$_1$ $_{-2}$* of mesh of three layers even starting from the Ferri 2 phase **[21].**

The rich polymorphism and the possibility of discovery of new chiral smectic phases have an extremely interesting. Several theoretical and models have been established to explain the variety of mesophases; their origin and structure, example are the HT model **[22,23] and its extension proposed by Dhaouadi et al [24].**

In this work we have carried out a theoretical study, based on the extended H-T model, of the field-induced states in chiral smectic liquid crystals, especially those obtained starting from SmC*$_{Fi2}$ phase. Two hypotheses of the dynamics that give rise to the intermediate phase have been discussed. The molecular arrangements within the layers of two suggested structures, with 3 and 4 layers, have been studied. We carried out a numerical calculation, aiming at comparing the free energies of the suggested structures, in order to know most stable between them.

Let us start initially with a theoretical background which describes the phase transitions between structures of the distorted clock model.

## B. Theoretical background

In this section we suggest to present briefly the phenomenological model, known as the Hamaneh Taylor model (H, T model). Thereafter we present the extension brought by *Dhaouadi et all* to this model. All the calculations carried out along this work are base on the results of this extension.

1. **The Hamaneh Taylor model**

The H-T model is a phenomenological model to describe chiral smectic phases [22,23]. It is based on the balance between two interactions; one a short range twisting term trying to impose an increment $\alpha$ of the azimuthal angle $\varphi$ between adjacent layers, the increment $\alpha$ of the azimuthal angle from layer to layer varies with temperature from 0 to $\pi$; the other a longue range related to the anisotropy of curvature energy in the layers plane. They derived an order parameter $J = <\cos(2\varphi_l)>$, where the average is taken on the azimuthal angle inside the unit cell. It is non-null in the commensurate phases and associated to an energy $\eta J^2$ where $\eta$ is a coefficient describing the strength of the longue range interaction and is of order unity. The short order term reads $<\cos(\Delta\varphi_l - \alpha)>$ with $\Delta\varphi_l = \varphi_{l+1} - \varphi_l$; where $\varphi_l$ correspond to azimuthal angle in layer l. So the free energy can be written as:

$$\frac{F}{F_0} = <cos\boldsymbol{\varphi_{l+1}} - \boldsymbol{\varphi_l} - \boldsymbol{\alpha}> + \eta J^2 \quad (1)$$

$F_0$ is the electrostatic energy necessary for transition to ferroelectric phase with a polarization $P_S$ at the electric field $E_s$. $F_0 = -P_s.E_s$.

From this expression Hamaneh and Taylor developed a numerical calculus leads to a phase diagram in the ($\alpha$, η) plane showing the sequence of sub phases which can be observed in a given liquid crystals.

**2. Extension of the H-T model**

An extension of the H-T model proposed by Dhaouadi et all **[24]** based on the introduction of new order parameter $I = <cos\varphi_l>$ which describe the contribution of macroscopic polarization and electric field. The contribution to the free energy of polarization and electric field are expressed respectively by

$$\Delta F_P = F_0 . \gamma . \sqrt{\eta} . I^2 \quad \text{and} \quad \Delta F_E = F_0 . \delta . \sqrt[4]{\eta} . I$$

Where γ and δ are two parameters having for order of magnitude the unit and have as expressions: $(\gamma = -\frac{P_S^2}{2 F_0 \varepsilon_0 \chi})$ and $(\delta = \sqrt{\frac{-2 \varepsilon_0 \chi}{F_0}} . E)$, $\chi$ is the electric susceptibility of the sample. So, the expression of the free energy is written:

$$\frac{F}{F_0} = \langle \cos(\Delta\phi_l - \alpha) \rangle + \eta J^2 + \gamma\sqrt{\eta}I^2 + \delta\sqrt[4]{\eta}I \quad (2)$$

From this expression, Dhaouadi et al derived new phase diagrams and correlated them to their experimental results showing the phase sequence in some compound as function of temperature and electric field.

We present in figure 2 three diagrams which show the evolution of the domains corresponding to the various phases of the liquid crystal for different values of the parameters γ and δ. The first, figure 2.a, is obtained starting from the H-T model without extension (γ = 0 and δ = 0). That of the figure 2.b is obtained starting from the extension of the H-T model at zero field and for (γ = 0.2). The dashed line shows one possibility for the form of the curve α (η). As we move down the curve, following the arrow, we pass from the SmC*$_{Fi2}$ to SmC* domains. Rising the temperature further takes us into the SmC$_\alpha$. Finally, η= 0 correspond to the SmA phase.

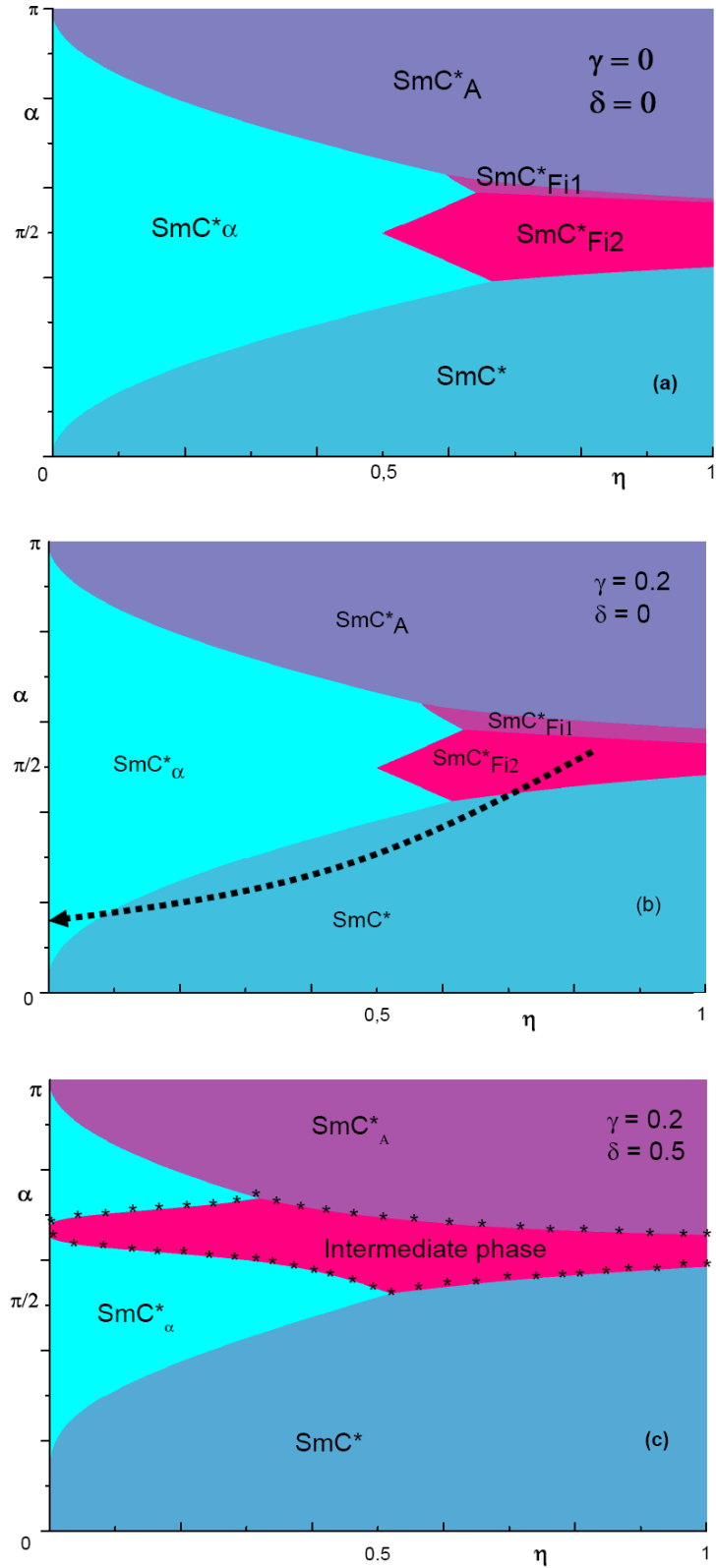

***Figure 2.*** *(α,η) phase diagrams obtained with; the H-T model (a); extended H-T model at zero field (b) and with applied field (c). Arrow shows evolution of curve α (η) on heating. The zone delimited by stars correspond to the domain where appears the intermediate phase.*

The diagram of figure 2.c is also obtained starting from the extension of the H-T model at non zero field ($\gamma = 0.2$ and $\delta = 0.5$).

To obtain these diagrams we compare the free energies of all the distorted clock model's known structures for different values of parameters ($\alpha$, $\eta$). The domains corresponding to the polar phases "SmC* and SmC*$_{Fi1}$" grow with the detriment of the other domains, when taking into account the polarization and as the electric field increases. It should be noted that for the diagram of the figure 2.c calculations are carried out by supposing that the intermediate phase has a structure with three layers similar to that of the SmC*$_{Fi1}$ phase.

The choice of the values of the parameters $\gamma = 0.2$ and $\delta = 0.5$ is made on the basis of experimental data obtained following an electro-optical study of the compound C12F3. A summary of experimental results is proposed in the following section.

## C- Experimental results

The aim of this section is to estimate the orders of magnitude of polarizations and the transition's threshold field which are crucial data for the interpretations of theoretical calculations carried out throughout this work.

An electro-optical study and microscopic observation have been carried out for the chiral smectic liquid crystal compound C12F3 belonging to the series of fluorinated product CnF3 synthesized by H.T Nguyen. **[5][25-27].** The structural formula of the compound C12F3 is drawn in figure 3.

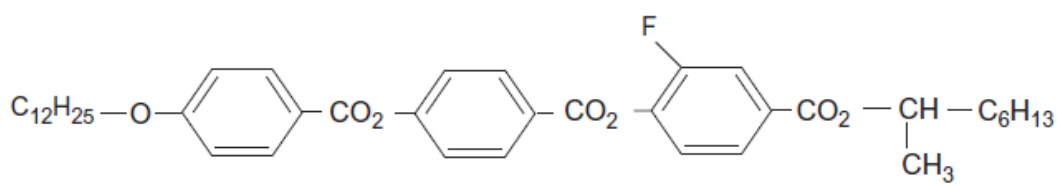

*The structural formula of the compound C12F3*

The polymorphism obtained by dielectric spectroscopy at zero fields is the following **[25]**:

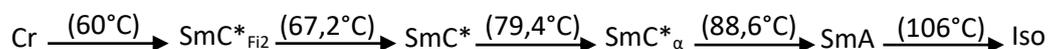

All measurements have been carried out on commercial planar cell (EHC. Inc, Japan; 5 µm) coated with a conductive layer of indium tin oxide. The active area is 25 mm².

1. **Microscopic observation**

We have recorded the evolution of the textures of our compound as a function of the applied electric field in the range of temperature where the sample presents the phase SmC *$_{Fi2}$. Photomicrographs of figure 4 are taken at the same temperature, **T=66.2°C.** For a weak field sample presents still the structure of the SmC*$_{Fi2}$ phase (fig 4.a). For stronger field, superior to the threshold value which one notes $E_{c1}$, the change of color means transition towards the intermediate phase (fig 4.b). A second transition towards the unrolled SmC* phase takes place for a field higher than a new threshold value noted $E_{c2}$ (fig 4.c).

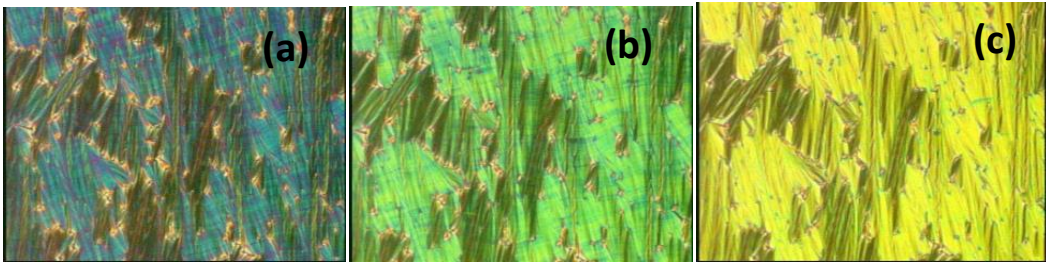

*Figure 3. Photomicrographs taken at T=66.2°C. (a) SmC*$_{Fi2}$ phase; (b) intermediate phase; (c) unwound SmC* phase.*

2. **Electro-optic study**

For the electro-optic study, we used a triangular wave form (see Figure 4). The measurement recorded at different temperatures, by varying voltages with fixed frequency 40 Hz in the 5-µm cell. We record the signal on a Tektronix 340 oscilloscope. The current response is the sum of the charge of the cell capacitor, the ionic conductivity and a *peak* linked to the polarization which reads:

$$i_p(t) = \iint \frac{\partial P}{\partial t}\, ds$$

and corresponds to the derivative versus time of the projection of the macroscopic polarization P over the direction y of the applied field. The measurement of the air of the peak makes it possible to determine the polarization of the sample

Figure 4 represents evolution of the switching current at T = 66.2°C for 4, 5 and 8 V (At 4V, a large current peak was observed, it correspond to the ferri 2/ intermediate state switching (fig.

4 a). A further increase of the voltage leads to a single current peak. It means that the switching occurs directly between the two ferroelectric states (fig. 4 b and c).

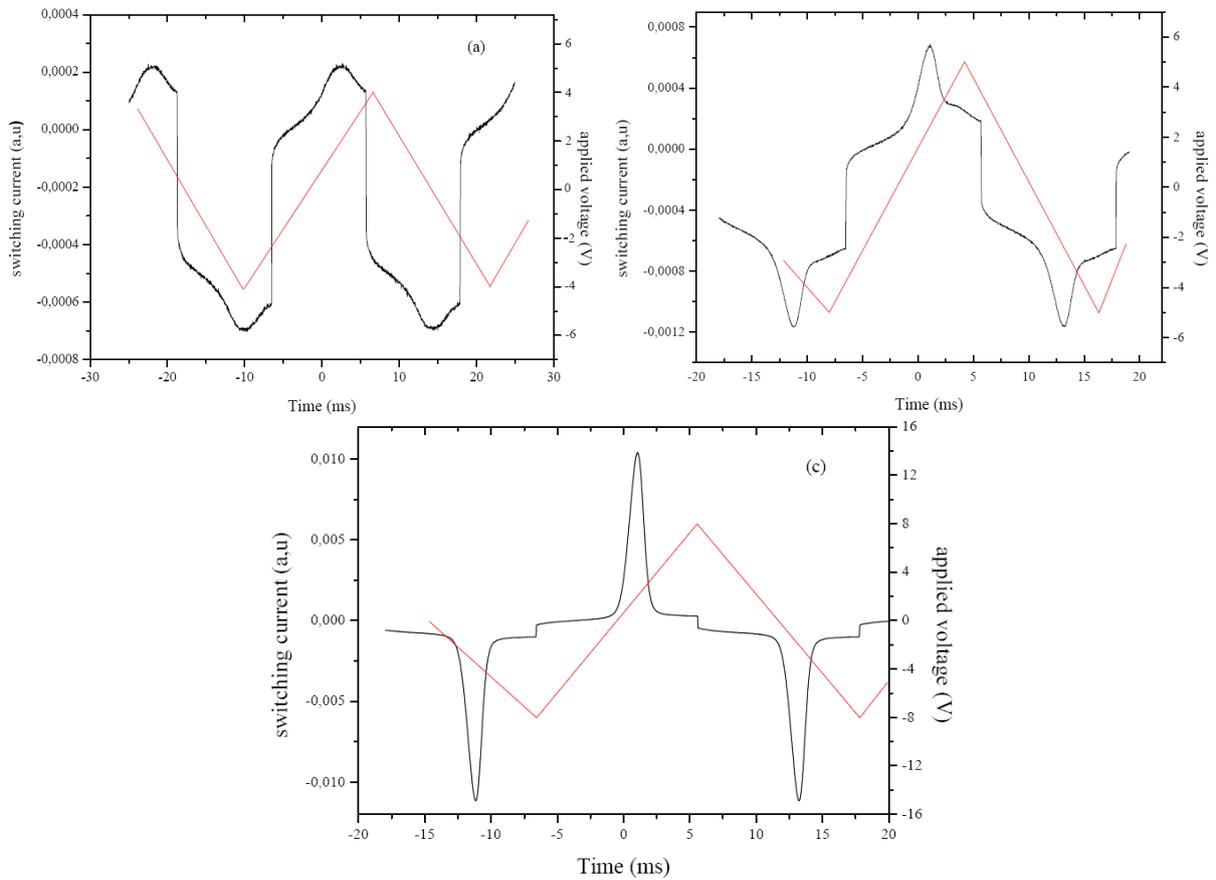

*Figure 4. Switching current for the SmC\*$_{Fi2}$ phase at 66.2°C, under a varying triangular electric field (frequency 40 Hz); (a) at 4V; (b) at 5V and (c) at 8V. The air of the peak is proportional to the polarization of the sample.*

The measurement of the polarization of the sample according to the applied field modulus at various temperatures enabled us to plot the curves of figure 5. The changes of shapes observed in the curve of the figure 5.a, indicate phases transitions. Thus, one determined the values of the threshold fields Ec and Es of the *SmC\*$_{Fi2}$* → intermediate phase and intermediate phase →*SmC\*$_{unrolled}$* transitions.

In figure *5b* we represent the evolution of the polarization of saturation Ps of the sample and the variation ΔP of polarization during the transition towards the intermediate phase in the range of temperature [62°C, 67°C]. The ratio ΔP/Ps is almost constant and equal to 2/5.

The figure *5c* represents the part of the (*E,T*) phase diagram of the compound *C12F3* relating to the range of temperature [62°C, 70.5°C]. The variation according to the temperature of the ratios *Es/Ec* and ΔP/Ps are represented in the figure *5d*. It is to be also noticed that the ratio *Es/Ec* is almost constant and equal to 1,4.

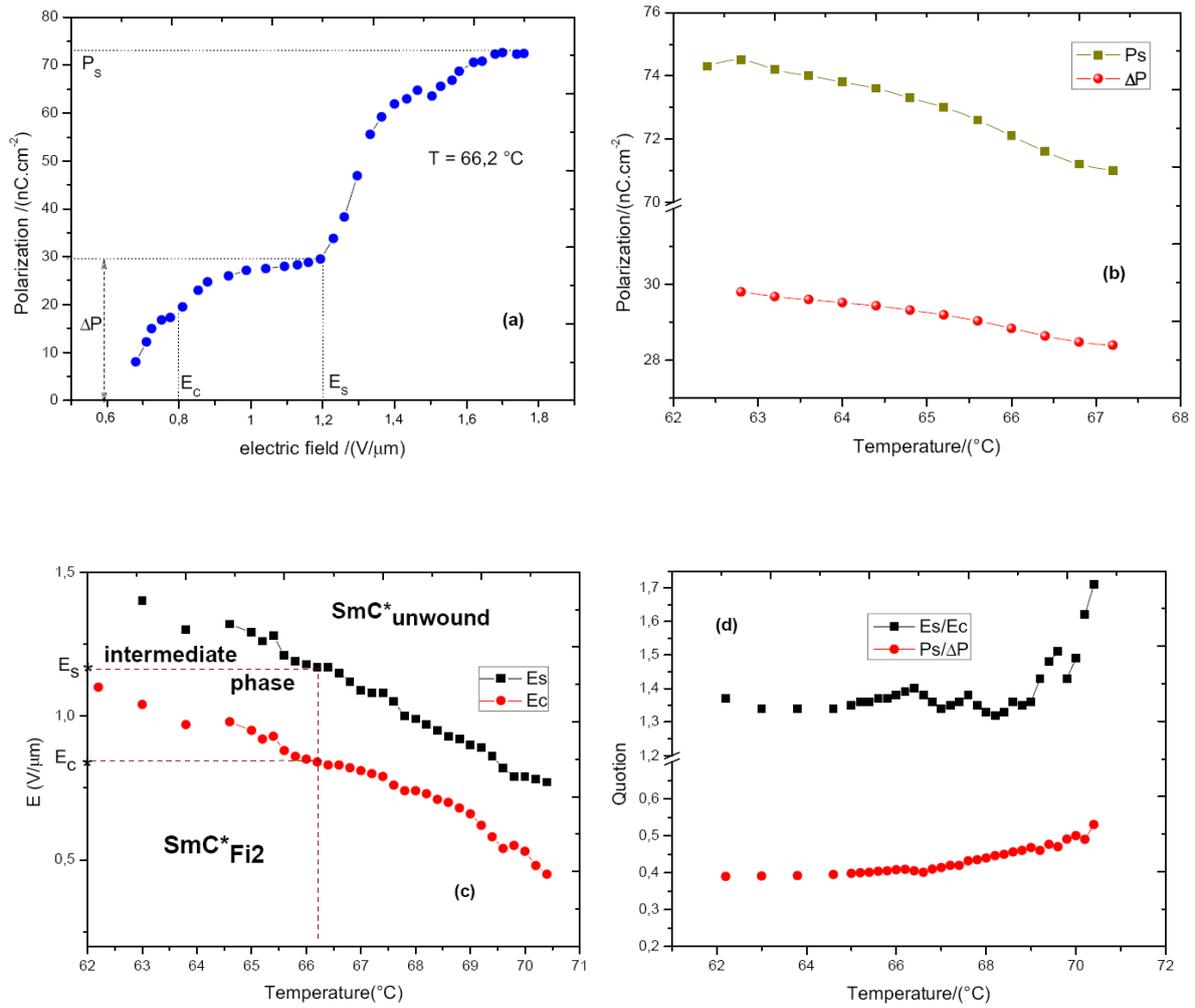

*Figure 5. Results of the electro-optical study ;(a) Polarization of the sample according to the applied field modulus at T = 66.2°C. The changes of shapes observed in the curve indicate phase transitions. (b) Evolution with temperature of the polarization of saturation Ps and the variation ΔP of polarization during the transition towards the intermediate. (c) (E,T) phase diagram of the compound C12F3. (d) Variation according to the temperature of the ratios Es/Ec and ΔP/Ps.*

## D. Origins of the field-induced intermediate phase

In this paragraph we will try to answer the following question: *which is the dynamics that gives rise to the intermediate phase?* To answer this question we propose a model according to which the applied electric field induces a distortion of the structure of the phase present in the sample at null field (figure 6.a). The study of the evolution of this distortion according to the applied field modulus will allow further explaining the appearance of this intermediate phase and deducing its possible structure.

1. **Working hypothesis**

The null field phase, to which this study relates, is the SmC*$_{Fi2}$ one. The fact that this phase is nonpolar facilitates enormously calculating of the distortions which deforms the structure following the application of an electric field. We suppose that the cell, into which the product is introduced, is low thickness so that the helix due to the chirality is destroyed by the planar anchoring imposed by treated surfaces. This assumption makes it possible to restrict the study to the elementary mesh of four layers.

Two assumptions describing the dynamics which leads to the transition towards the intermediate phase are considered in what follows:

- The first is that of a deformation of the basic structure with conservation of the initial four layers mesh. The transition towards the intermediate phase is explained by a discontinuity of this distortion (figure 6.b).
- The second consists on a total rearrangement of the initial mesh. The basic four layers structure will be destroyed when the electric field exceeds the threshold value $E_{c1}$. A new structure with three layers, identical to that described by S. *Jaradet* [20], appears in its place (figure 6.c).

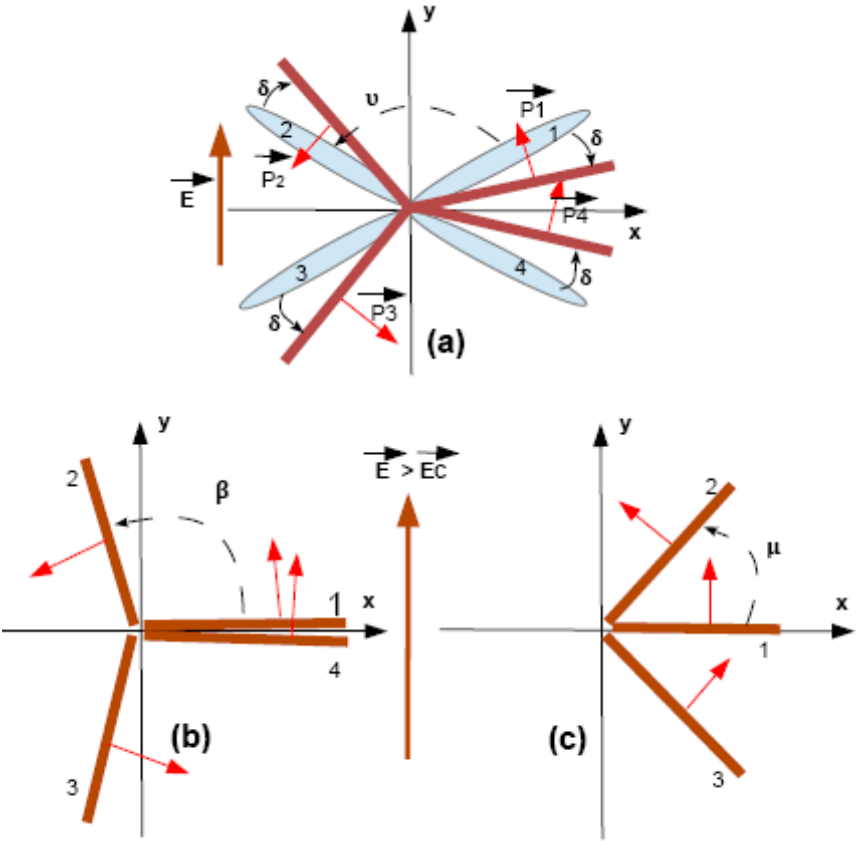

*Figure 6. (a) Distortion of the four layer structure for E < $E_c$. (b) First assumption with discontinuity of the distortion for $E_c$ <E < $E_s$. (c) Second assumption appearance of a new structure with three layers for $E_c$ <E < $E_s$. Red arrows represent the directions of polarization in the various layers.*

Initially we will carry out a calculation of polarization in each of the two structures described on the assumptions below. Then, a comparison of their free energies makes it possible to determine which the most probable one is.

## 2. Distortion of the elementary mesh in the *SmC* *$_{Fi2}$ phase

(Figure 6. a) show the evolution of molecular arrangements in the elementary mesh of the *SmC**$_{Fi2}$ phase, following the application of an electric field. The distortion of the basic structure appears as an angular displacement «$\delta_i$» of the molecular orientation in layer "i" (with i= 1, 2, 3). For the simplicity of calculations, one supposes that all the displacement «$\delta_i$» are identical. The resulting structure presents a spontaneous polarization due to the new orientations of molecular dipole moments. A calculation of the distortion «$\delta$» and induced polarization «*P*» are necessary for the study of the transition towards the intermediate phase.

- **expression of the free energy**

In the expression of the free energy of the H-T model (*equation.1*), it is the second term corresponding to the long range interactions $[\![F_0\, \eta\, J^2]\!]$ which is responsible for the appearance of the commensurable molecular arrangements **[24].** It is also responsible for the stability of the distorted structure described previously.

We note *W* the potential governing the stability of these structures. In the *SmC**$_{Fi2}$ phase, the order parameter J is written at null field as:

$$J = <\cos(2\,\phi_l)> = \cos(\upsilon) \quad (3)$$

*W* is minimal for the structure described in the *figure.5.a*. ($W = W_0$ for $\delta = 0$).

$$W = F_0\,\eta\,J^2 \text{ and } W_0 = F_0\,\eta\,\cos^2(\upsilon) \quad (4)$$

In the structure distorted by the applied electric field, the order parameter J and the potential $W_0$ are a function of the distortion «$\delta$».

$$J = \frac{1}{4}\left(\cos 2(-\frac{\upsilon}{2}-\delta) + \cos 2(\frac{\upsilon}{2}-\delta) + \cos 2(\pi - \frac{\upsilon}{2}+\delta) + \cos 2(\pi + \frac{\upsilon}{2}+\delta)\right) \quad (5)$$

$$W(\delta) = \eta\,F_0\,\cos^2(\upsilon)\,\cos^2(2\delta) \quad (6)$$

A development to the order "2", for low values of the distortion «$\delta$», gives for the potential the following expression:

$$W(\delta) = \eta\, F_0 \cos^2(\upsilon) - 4\, \eta\, F_0 \cos^2(\upsilon).\delta^2 = W_0 - 4\, W_0\, \delta^2 \qquad (7)$$

The expression of the free energy must present in addition to the potential $W(\delta)$, a term corresponding to the energy of the polarization which appears at non null field. For a linear treatment one neglects the quadratic term in $P^2$. A justification of this approximation will be given further. The symmetry of the configuration of the figure *5.a* gives an induced polarization parallel to the applied field.

$$F^* = W(\delta) - P(\delta).E \qquad (8)$$

- **Expression of the distortion «$\delta$», and polarization «$P$»**

Let us start initially by expressing polarization $P(\delta)$ according to $\delta$. This polarization is the average of all polarizations in the mesh layers after distortion.

$$<P> = 1/4\,[P_1 + P_2 + P_3 + P_4] \qquad (9)$$

The components of polarizations $P_1$, $P_2$, $P_3$ and $P_4$ following the direction of the field are written:

$$P_{S1} = P_S \cos\left(\frac{\pi}{2} - \left(\frac{\upsilon}{2} + \delta\right)\right), \quad P_{S2} = -P_S \cos\left(\frac{\pi}{2} - \left(\frac{\upsilon}{2} - \delta\right)\right),$$

$$P_{S3} = -P_S \cos\left(\frac{\pi}{2} - \left(\frac{\upsilon}{2} - \delta\right)\right), \quad P_{S4} = P_S \cos\left(\frac{\pi}{2} - \left(\frac{\upsilon}{2} + \delta\right)\right) \qquad (10)$$

Total polarization is written:

$$P = P_s \cos\left(\frac{\upsilon}{2}\right) \sin(\delta) \qquad (11)$$

For weak values of $\delta$, $\sin(\delta) \approx \delta$ one has then:

$$P(\delta) = P_s \cos\left(\frac{\upsilon}{2}\right) \delta \qquad (12)$$

The free energy is then written in the following form:

$$F^* = W(\delta) - P(\delta).E = W_0 - 4.W_0.\delta^2 - Ps \cos\left(\frac{\upsilon}{2}\right).E.\delta \qquad (13)$$

By minimization over δ one finds:

$$\delta = \frac{-P_s \cos\left(\frac{v}{2}\right) E}{8 W_0} = \frac{\cos\left(\frac{v}{2}\right)}{8 \eta \cos(v)^2} \frac{E}{E_s} \quad (14)$$

Where $P_s$ and $E_s$ are respectively the polarization of saturation of the sample and the field threshold which induces the transition towards the unrolled SmC* phase.

### 3. Order of magnitude

To estimate the order of magnitude of the angular distortion and polarization induced by this distortion it is necessary to know the order of magnitude of the characteristic angle $v$ of the *SmC\*$_{Fi2}$* phase. Starting from resonant x-ray scattering measurements, Cady *et al.* **[28]** found for the Sm*C\*$_{Fi2}$* structure a value of $v$ of about 164°. Roberts *et al.* **[29]** measured the angular distortion of the Sm*C\*$_{Fi2}$* structure in mixtures at two temperatures; they found a value of $v$ of about 166° with no discernable dependence on temperature. We adopt in what follows the value ($v = 164°$).

- **The distortion and the induced polarization**

According to the H-T model, the *SmC\*$_{Fi2}$* phase is obtained for values of $\eta$ higher than 0,5. It is then possible to calculate the value of the distortion $\delta$:

$$\delta = \frac{0{,}13}{8 \cdot 0{,}5 \cdot 0{,}933} \frac{E}{E_s} = 0{,}034 \frac{E}{E_s} \quad rd \quad (15)$$

Figure 5. d show that in a large range of temperature, the transition towards the intermediate phase occurs for the critical field $E_C$ such as the ratio $\frac{E_S}{E_C} \approx 1{,}4$, where $E_S$ is the threshold field inducing the transition towards the *SmC\*$_{unrolled}$* state. The equation (15) gives then a maximum value of the distortion $\delta$ of about de $\delta_{max} \approx \frac{0{,}034}{1.4}$ rd = 0,024 rd ≈ 1.37°.

While replacing $\delta$ by his value in the equation (12), one finds the value of the polarization which the sample following this process of distortion can have: $P_{max}(\delta) = 0{,}0031\ P_s$. It is to be noticed that even for rather strong fields ( $E \approx 0.7\ E_s$) the values of the distortion $\delta$ and polarization P remain low in front of the characteristic angle $v$ and polarization of saturation $P_S$ of the *SmC\*$_{Fi2}$* Phase. To conclude, a simple distortion of the structure *SmC\*$_{Fi2}$* cannot explain the appearance of the intermediate phase with a polarization close to half of $P_S$.

- **Discontinuity of the distortion; The four layers Structure**

We initially consider the four layers structure described in the figure 6 b. the transition towards the intermediate phase results in a discontinuity of the distortion « δ ». The average polarization of the new structure is written according to the angle β :

$$\langle P \rangle = \frac{1}{4}(P_{S1} + P_{S2} + P_{S3} + P_{S4}) = \frac{P_s}{2}(1 + \cos(\beta)) \quad (16)$$

Evolution of the polarization of the sample according to the applied electric field at the temperature T = 66.2 (figure 5.d) shows that the polarization of the intermediate phase is about $\frac{2}{5}$ P$_s$. Starting from equation 16, the calculation of the new characteristic angle β gives; β ≈ 101°, which corresponds to a value of the distortion $\delta_0$ of order 71.

- **The three layers structure**

This new structure is similar to that observed in experiments by S. Jaradet **[20]** (figure 1). We carry out in the continuation a calculation of the characteristic angle "μ" of this new phase.

According to the structure suggested in the figure 6 c, one expresses the variation of polarization ΔP as function of the angular parameter $\mu$ and the polarization of saturation P$_S$.

$$\Delta P = \frac{1}{3}[P_{S1} + P_{S2} + P_{S3}] = \frac{1}{3}P_S[1 + 2\cos\mu] \quad (17)$$

To estimate the order of magnitude of the angle $\mu$, one takes the value of $\Delta P = \frac{2}{5}$P$_s$ mentioned previously. One finds a value of $\mu$ of about 84,26°.

## E. modélisation numérique et diagramme de phases

In what follows we carry out a numerical calculation allowing a comparison of the free energies of the two suggested structures of the intermediate phase. A *Maple* program was elaborated to carry out this task.

The expression of the free energy (*équation.2*), written for these two structures, is used to plot the phase diagrams in the plan (*α, η*). The program ensures a numerical calculation which compares the absolute minima of the free energies associated to the two structures and indicates which is most stable. The program creates iterative loops to repeat successive minimizations with different values of *α* and *η* at each iteration. The construction of the diagrams is carried out by an interface to create graphs starting from matrices containing $10^6$ elements.

For a couple of selected values of the characteristic angles $\beta$ and µ, the obtained phase diagram shows the domains of stability of the two structures suggested for the intermediate phase.

The diagrams 7 a, 7 b, 7 c and 7 d correspond to choices of the angles $\beta$ and µ with values equal or slightly different from the optimal values $\beta = 101°$ and µ= 84.26° reported in the section order of magnitude. The thin blue band in bottom of these diagrams corresponds to a domain where the free energy of the four layers structure is weaker than that of three layers.

Even with a choice of the couple ($\beta$,µ) very different from the optimal values, the blue band changes thickness but the domain of stability of the three layers structure always remains dominating figure 7 e and 7 f.

The first conclusion to be made, according to the diagrams of figure 7, is that the three layers phase is the most stable in the zone of the plan ($\alpha$, $\eta$) delimited by stars, zone corresponding to the domain where appears the intermediate phase in diagrams H-T of the figure *2.c*.

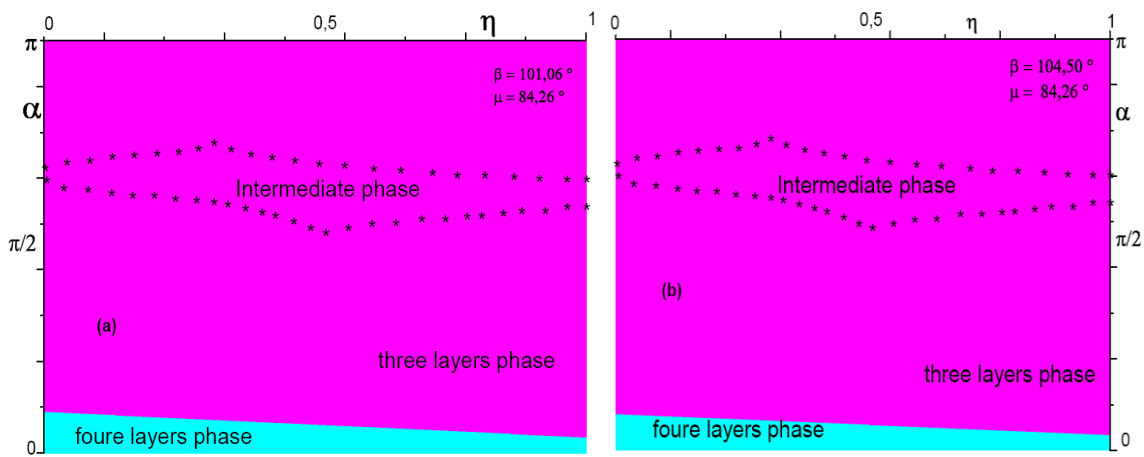

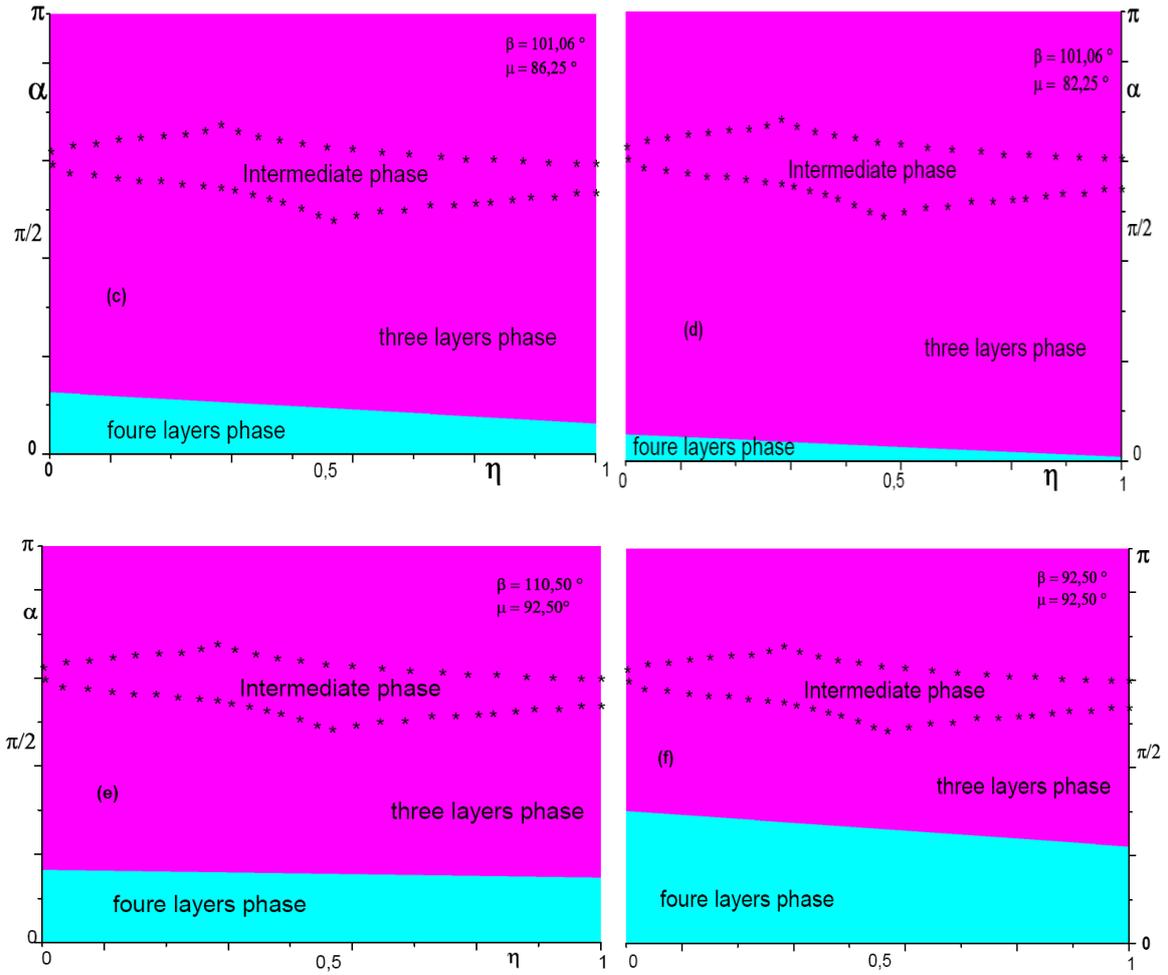

*Figure 7. (α, η) phase Diagrams drawn up numerically with different values of the parameters µ and β. They show the domains of stability of the two suggested structures for the intermediate phase. In blue the domain of four layers phase and in pink that of the three layers one. the zone delimited by stars correspond to the domain where appears the intermediate phase in diagrams H-T of the figure 2.c.*

Finally the results of numerical calculation confirm the experimental results according to which the intermediate phase admits a structure with 3 layers.

## Conclusion

Following the application of an electric field, a chiral smectite liquid crystal undergoes modifications of structure. Certain modifications constitute true phase transitions similar to those induced by variation in the temperature **[30].** It is well-known for a long time that under an electric field, chiral smectic liquid crystals transit usually to the unwound SmC* phase where the helical structure is completely unrolled. Sometimes the sample transits initially towards an intermediate polar state, (ferrielectric), before the total destruction of the helix.

One can thought that we are observing a progressive and discontinuous destruction of the helix which occurs by increasing the modulus of the applied field **[15].** Actually they are well true transitions from first order with coexistence of the two concerned phases, figure3.

In this work we tried, using a simple theoretical study, to understand the process which gives rise to this intermediate phase in the case of a sample presenting the *SmC* *$_{Fi2}$ phase at null field. Two possible structures of this phase are proposed.

   - The first is a four layers structure, which rises from a discontinuity of the distortion of the basic structure.

   - The second is a three layers structure, similar to those observed in experiments by the technique of X-ray resonant diffraction.

A theoretical study allows, starting from experimental data, to estimate the characteristic parameters of the suggested structures. A numerical calculation allows the comparison of their free energies to determine which of both is most stable. The results of this numerical study give favor to the three layers structure.